\documentclass[sigconf]{acmart}

\renewcommand\footnotetextcopyrightpermission[1]{}

\usepackage{xcolor}
\usepackage{subfig}
\usepackage{amsmath}
\usepackage{physics}
\usepackage{booktabs}
\usepackage{graphicx}
\usepackage{textcomp}
\usepackage{algorithmic}
\usepackage{filecontents}
\newcommand{\sol}{Qompose}
\newcommand*\mean[1]{\bar{#1}}

\begin{document}

\title{\sol{}: A Technique to Select Optimal Algorithm- Specific Layout for Neutral Atom Quantum Architectures}

\begin{abstract}
As quantum computing architecture matures, it is important to investigate new technologies that lend unique advantages.  In this work, we propose, \sol{}, a neutral atom quantum computing framework for efficiently composing quantum circuits on 2-D topologies of neutral atoms. \sol{} selects an efficient topology for any given circuit in order to optimize for length of execution through efficient parallelism and for overall fidelity. 
our extensive evaluation demonstrates the \sol{} is effective for a large collection of randomly-generated quantum circuits and a range of real-world benchmarks including VQE, ISING, and QAOA.\vspace{-2mm}
\end{abstract}

\author{Daniel Silver}
\affiliation{%
  \institution{Northeastern University}
  \city{Boston}
  \country{USA}}
  
\author{Tirthak Patel}
\affiliation{%
  \institution{Rice University}
  \city{Houston}
  \country{USA}}
  
\author{Devesh Tiwari}
\affiliation{%
  \institution{Northeastern University}
  \city{Boston}
  \country{USA}}
  
\maketitle

\section{Introduction}

Quantum computing architectures are evolving rapidly -- with multiple potential technologies (e.g., superconducting, ion-trap, neutral-atom, and photonics-based quantum computing) offering different design trade-offs \cite{patel2022optic, silver2023sliq, kielpinski2002architecture, henriet2020quantum, ranjan2024proximl}. Neutral-atom based quantum architecture is among the most promising and emerging quantum computing technology -- compared to traditional 
superconducting and ion-trap technologies,  neutral-atom based quantum architecture offers three key advantages: (a) lower cooling requirements, (2) longer coherence times and better scalability, and (3) flexibility to arrange the basic unit of computation (qubits) in an algorithm-specific manner. Physicists and material scientists are making significant strides toward maximizing the first two key advantages~\cite{picken2018entanglement,malbrunot2022simulation} including the most notable experimental demonstrations using neutral atoms~\cite{graham2022multi}. While the neutral atom hardware technology is maturing, its third benefit has received limited attention from the computer systems and architecture community. 

In particular, it has been shown recently that neutral atoms can be arranged in different topologies and the quantum circuit can be mapped on the chosen topology to execute the quantum operations in the circuit~\cite{saffman2016quantum}. Unfortunately, there is a limited understanding of this flexibility of atom arrangement that can be leveraged to execute different quantum algorithms more efficiently --- for example, produce better answer quality in the presence of error, reduce the number of operations (i.e., physical laser pulses), etc. To explore this space, we performed a series of experiments. Our results (discussed in detail in Sec.~\ref{sec:motiv}) reveal that arrangement of atoms (topology) significantly affects the execution of different quantum algorithms. Different quantum algorithms may prefer different topologies of atoms -- and, notably, as we show, these topologies are not necessarily very different or complex from each other, but their impact on the overall program execution is significant. Second, we also discovered that the preferred topology is also dependent on the figure of merit the user/system is trying to optimize (e.g., answer fidelity, pulse count, etc.) -- the choice of optimization metric depends on various algorithms and system environment constraints in neutral atom-based quantum computing~\cite{saffman2016quantum}.

Therefore, motivated by these experimental observations, the goal of this work is to demonstrate how different practically feasible and simple arrangements of neutral atoms can be leveraged to improve the overall execution of quantum circuits in an algorithm-specific way. However, we show, that this problem poses non-trivial challenges due to the inherent complexities of the neutral atom-based quantum computing architecture and execution of quantum circuits. One challenge is selecting a topology from the infinite space of possible topologies.  Even when we scope the space to three practical topologies, (square, s-triangle, and t-traingle; details in Sec.~\ref{sec:background}), each topology offers its own unique advantages and disadvantages. However, selecting the most suitable topology among three related but unique topologies is a challenging task because it is not possible to know apriori the critical execution path of a quantum circuit and the degree of operation parallelism a quantum circuit can exploit when it mapped onto a specific topology.


Therefore, we build a learning-based solution that determines the most effective topology of atoms for a given quantum algorithm. Some of the features used are quantum-specific (e.g., entanglement variance) and some are classical (e.g., page rank standard deviation). By leveraging a neural network, \sol{} combines these quantum and classical features into a prediction model that can anticipate the best topology for a given circuit.

\vspace{1mm} 

\noindent\textbf{Contributions.} This paper makes the following contributions:

\vspace{1mm} 

\noindent\textbf{I.} \textit{First, to the best of our knowledge, this is the first study to demonstrate that, under neutral atom-based quantum computing architecture, the arrangement of atoms (topology) significantly affects the execution and outcome of different quantum algorithms.} A single topology (e.g., a traditionally used square topology of atoms) does not necessarily produce the highest answer quality for all quantum algorithms -- in fact, even simple but slightly different topologies can produce significantly different answer qualities. 

\vspace{2mm} 

\noindent\textbf{II.} \textit{Second, \sol{} is the first solution that exploits this opportunity space and builds a learning-based solution that determines the most suitable topology of atoms for a given quantum algorithm from a set of practically-feasible and simple topologies.} \sol{} builds a new methodology, inspired by the page-rank algorithm \cite{page1999pagerank}, that identifies and encodes the properties of quantum circuits that are most critical toward learning and predicting the most suitable topology.

\vspace{2mm} 

\noindent\textbf{III.} \textit{Finally, our evaluation demonstrates the \sol{} is effective in selecting the best neutral-atom topology -- for a large collection of randomly-generated quantum circuits and a range of real-world benchmarks including VQE, ISING, and QAOA}. In particular, \sol{} improves the pulse counts in the critical path by 8.4\% and 5.4\% for randomly-generated quantum circuits and real-world benchmarks over randomly selecting one of the topology options. Our extensive evaluation shows that \sol{} almost always (except in one of 19) selects the most-effective topology for a given figure of merit and benchmark.
\vspace{0.0mm} 

\noindent\textit{\underline{Our open-source artifact} is the first experimental framework that enables researchers to explore different neutral-atom topologies, quantify their trade-offs, and optimize the topologies in an algorithmic-specific manner:} \url{https://anonymous.4open.science/r/Qompose-5B67/}

\section{Background}
\label{sec:background}

We organize this section into two subsections. We first begin by discussing general information regarding quantum computing, then discuss neutral atom quantum computing in more detail.

\subsection{Quantum Computing Background}

\subsubsection{Quantum States}

Analogous to the ``bit'' of the conventional computing world, which we will hereon refer to as classical computing, the quantum computing world has a building block known as a ``qubit''. A qubit is much more flexible then a bit as while a bit can only exist as a 1 or 0, a qubit may exist as a superposition of these states. This can be well-represented using Dirac notation $\ket{\Psi} = \alpha\ket{0} + \beta\ket{1}$ where $\alpha$ and $\beta$ are complex coefficients bound by $\norm{\alpha}^2 + \norm{\beta}^2 = 1$ \cite{textbook}. Similarly to how the general equation of a circle centered around the origin is defined as $x^2 + y^2 = r^2$, $\norm{\alpha}^2 + \norm{\beta}^2 = 1$ can be expanded to be visualized as a point on a sphere with radius $1$, known as the Bloch Sphere \cite{patel2021qraft}. Additionally, while qubits may exist in a superposition, once they are measured, they probabilistic collapse to a classical state is defined by the complex coefficients of the states.

\subsubsection{Quantum Circuits}

Quantum gates can be used to perform logical operations on qubit states. When many quantum gates are placed in sequence to perform desired logical operations, they are referred to as a quantum circuit. While any classical gate has a logical quantum equivalent gate, it must be in the form of a unitary matrix satisfying the condition $U^\dagger U = I$, where $U^\dagger$ where $U^\dagger$ is defined as the complex conjugate transpose of $U$. Notably, this ensures full reversibility of all quantum circuits. These gates can be built for any number of qubits. An example of a 3-qubit quantum gate is the $CCX$ gate: defined for 3 qubits $a$, $b$, and $c$:
$CCX_{a,b,c} = \ket{0} \bra{0} \otimes I \otimes I + \ket{1} \bra{1} \otimes CX$, where $CX$ is a two-qubit gate \cite{Qiskit}.


\subsection{Neutral Atom Quantum Computing}

As practical quantum computing is only at its infancy, it is unclear which technologies will reign as the best standard \cite{baker2021exploiting}. Some of the many technologies currently being explored are superconducting qubits, ion traps, photonics, and neutral atom quantum computing \cite{Qiskit}.  Neutral atom quantum computing comes with many unique qualities that make it a promising technology for quantum computation. Specifically, neutral atom quantum computing allows for arbitrary arrangement of 2D topologies for qubit positioning \cite{saffman2016quantum}, enabling incredible parallelizability of operations. In these systems, atoms are trapped through optical tweezer arrays, which can then be arranged into any shape where one atom acts as one qubit. The qubits states are based on the energy levels of the valence electrons (typically using Rubidium and Cesium atoms). These arrays have been shown to be loaded on a 100ms timescales on 1D \cite{endres2016cold} and 2D \cite{barredo2016atom} topologies using the aforementioned tweezer method.

\begin{figure}[t]
	\centering
    \includegraphics[scale=.3]{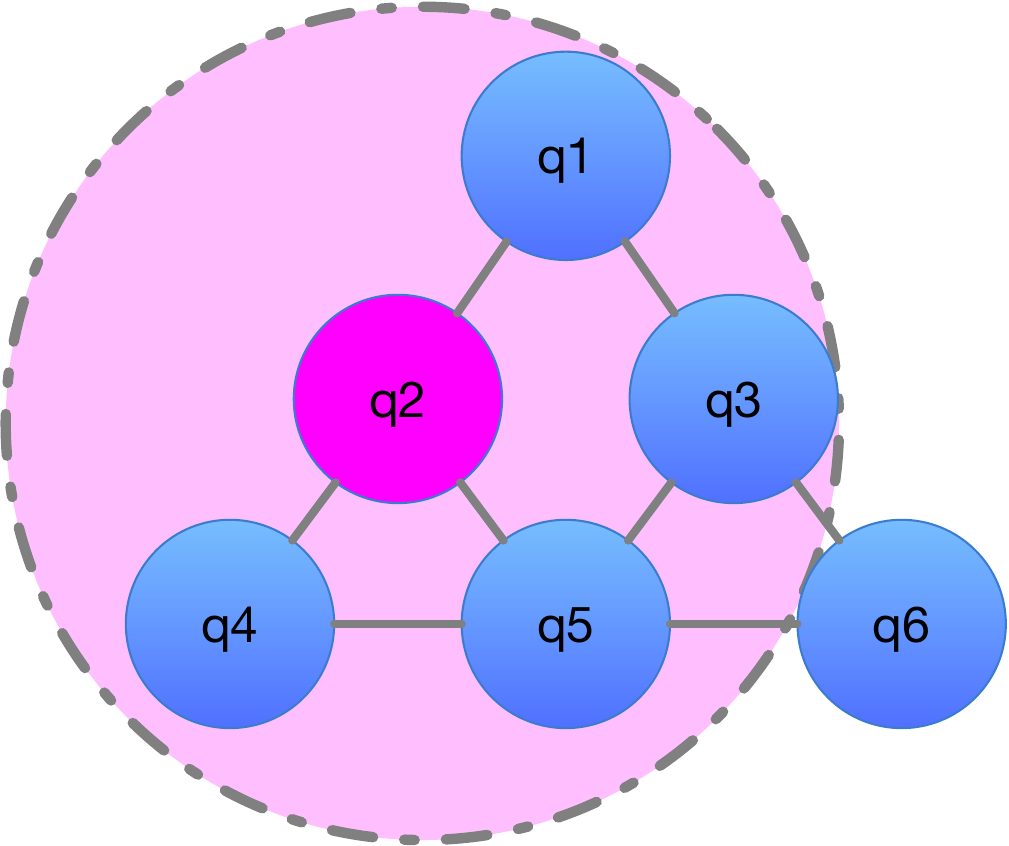}
    \vspace{1mm}
    \hrule
    \vspace{-3mm}
	\caption{As a Rydberg interaction is created around qubit q2, all adjacent qubits are blocked in the restriction zone and cannot be entangled with any other qubits using the same frequency at the same time.}
	\label{restriction}
	\vspace{-4mm}
\end{figure}

\begin{figure}[t]
	\centering
    \includegraphics[scale=0.2]{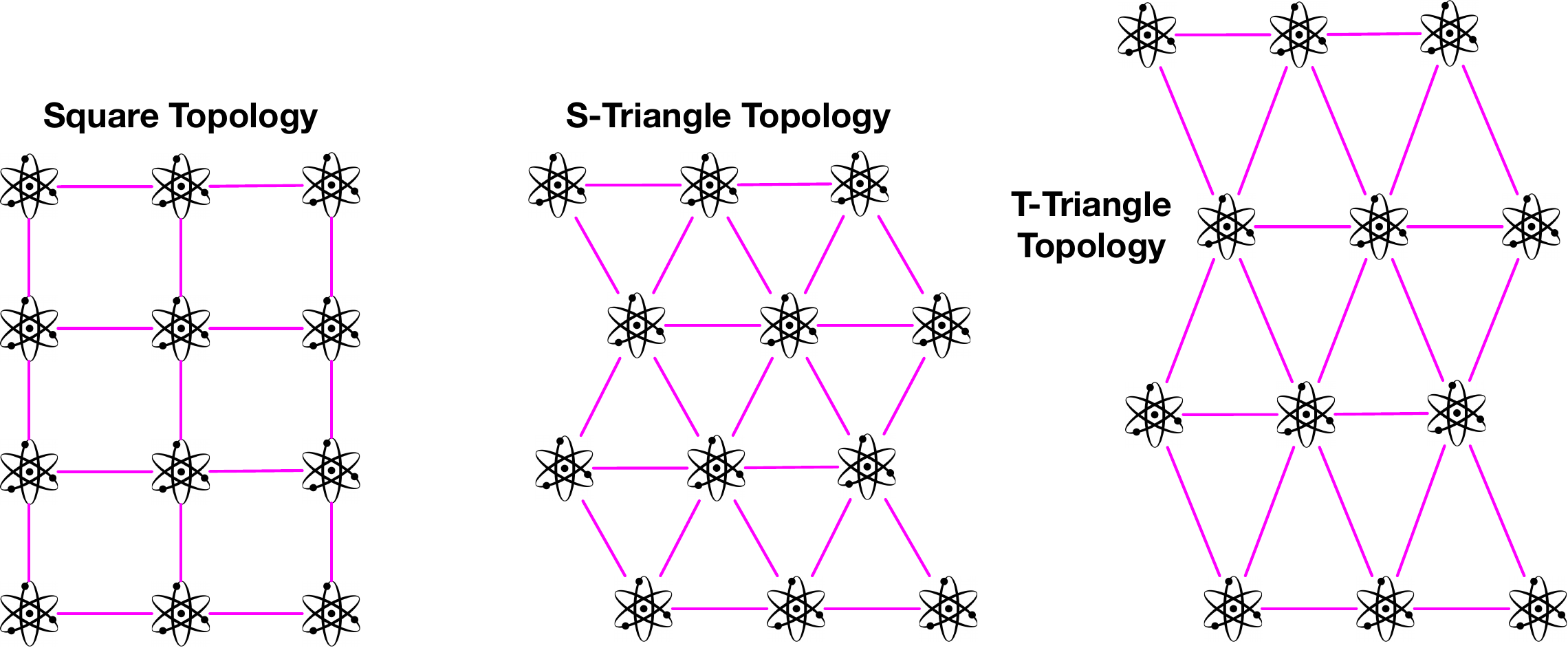}
    \vspace{1mm}
    \hrule
    \vspace{-3mm}
	\caption{Topology arrangement options. The three topologies pictured each pose unique advantages for certain circuit types. For example, lower circuit width accurately predicts the s-triangular topology for reducing the critical pulse path.}
	\label{topos}
	\vspace{-4mm}
\end{figure}

\subsubsection{Neutral Atom Manipulation}

Different light pulses are used to excite electrons into states. Single qubit gates are excited using a Raman transition, requiring a single light pulse. Multi qubit gates require Rydberg transitions, which require multiple pulses and are used to perform actions between neighbors, while entangling them with one another.

\subsubsection{Restriction Zones}

Rydberg interactions occur within a given radius and affect all atoms within the radius. These interactions entangle all qubits within the radius and effectively block any other operation using the same frequencies from taking place on those qubits at the same time. While these qubits are blocked however, any operation outside of the restriction zone can have the same pulses applied in parallel. Additionally, qubits undergoing operations of the same frequency within this zone are unaffected. For example, a particular qubit may be in a single qubit Raman interaction at the same time as it is undergoing a Rydberg interaction.  This restriction zone behavior is visualized in  Figure \ref{restriction}.

\section{Motivation for \sol{}}
\label{sec:motiv}

Quantum computing has the potential to enable exponential speedups over classical algorithms in many problem domains. To enable this, neutral atom quantum computing has many beneficial properties such as topological flexibility, operating at warmer temperatures \cite{weiss2017quantum}, and long distance interactions \cite{baker2021exploiting}, which could make it the most reasonable technology for many quantum computing tasks. 

\textbf{Specifically the ability to arrange the qubits in arbitrary positions can enable very high degrees of parallelism}. By making circuits more parallel, the error rate can be lowered as the it compounds less along the critical path. Additionally, the pulse count along the critical path can vastly improve the execution time of the circuit and free up valuable resources for more circuits. The total pulse count, which aggregates the pulses of every gate operation, affects the overall error rate that an algorithm may experience. We demonstrate how unique topologies can create vastly different outcomes for each of these important measurements along a compiled circuit.  

In this work, we pose the question of \textit{how to leverage topological flexibility to optimizing general quantum circuits}. We find that the distinction of topology can drastically change the execution of the circuit and therefore improve or degrade performance. 

\begin{figure}[t]
	\centering
    \includegraphics[scale=0.194]{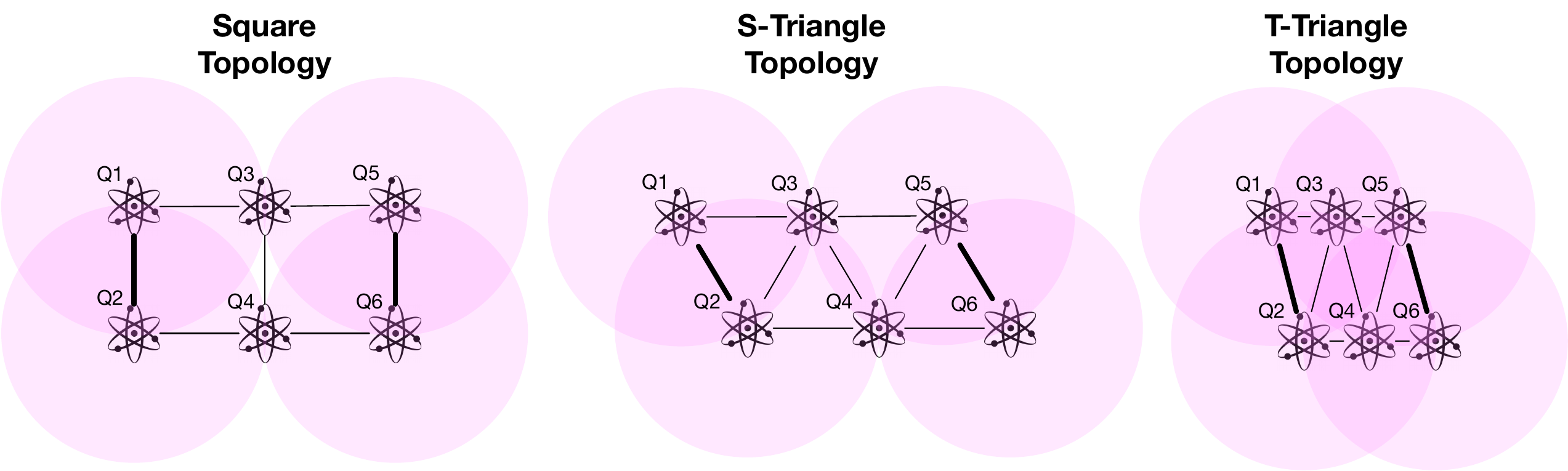}
    \vspace{1mm}
    \hrule
    \vspace{-3mm}
	\caption{Demonstration of activation radius blocking. Each topology blocks nearby qubits differently, which is what leads to unique advantages offered by each.}
	\label{compare}
	\vspace{-4mm}
\end{figure}

\begin{table}[t]
\caption{Critical Pulse Count Comparison}
\vspace{-2mm}
\begin{tabular}{||c c c c||} 
 \hline
 Benchmark & 1st & 2nd & 3rd \\ [0.5ex] 
 \hline\hline
 HLF & S-Triangle & Square & T-Triangle \\ 
 \hline
 Mult25 & Square & T-Triangle & S-Triangle \\
 \hline
 QFT15 & T-Triangle & S-Triangle & Square \\
 \hline
\end{tabular}
\label{tab:difalgo}
\end{table}

\begin{table}[t]
\caption{Mod5D1 Benchmark Topology Evaluation}
\vspace{-2mm}
\begin{tabular}{||c c c c||} 
 \hline
 Metric & 1st & 2nd & 3rd \\ [0.5ex] 
 \hline\hline
 Fidelity & T-Triangle & S-Triangle & Square \\ 
 \hline
 Critical Pulses & S-Triangle & Square & T-Triangle \\
 \hline
 Total Pulses & S-Triangle & T-Triangle (Tie) & Square (Tie) \\
 \hline
\end{tabular}
\label{tab:samealgo}
\end{table}

\begin{figure*}[t]
	\centering
    \includegraphics[scale=0.39]{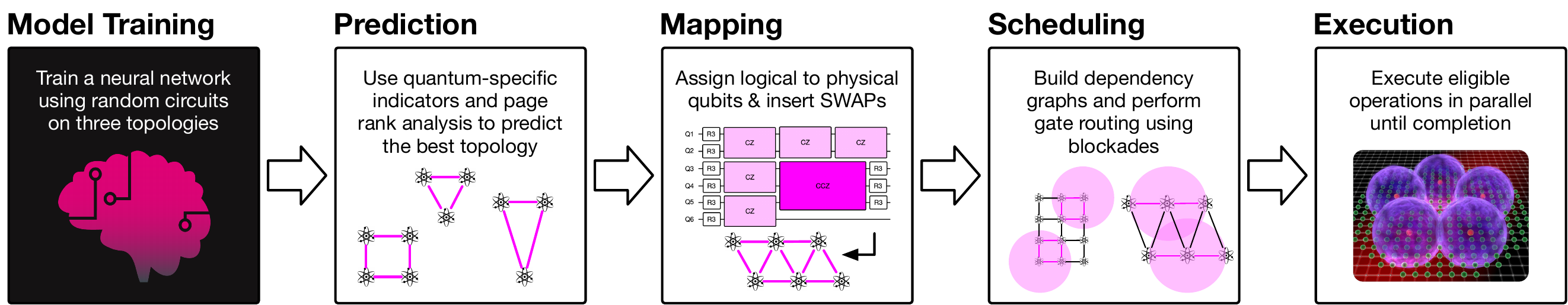}
    \vspace{1mm}
    \hrule
    \vspace{-3mm}
	\caption{An overview of all of the steps of \sol{}: model training, topology prediction, mapping, scheduling, and execution.}
	\label{overview}
	\vspace{-4mm}
\end{figure*}

Three different topologies of neutral atom arrangement are shown here and are present throughout this work: Square, S-Triangle, and T-Triangle (as shown in Fig.~\ref{topos}). Square is a commonly used grid \cite{baker2021exploiting}, while the other two topologies are novel and were found to work efficiently because of the trade-offs that they provide. These trade-offs affect the impact of different topologies on circuit performance is due to the differences in how operations are performed. For example, Fig.~\ref{compare} shows the scheduling of two $CX$ gates on qubits Q1-Q2 and Q5-Q6. They can be run in parallel on the Square and S-Triangle topologies as there are no overlapping blockades. However, they cannot be run in parallel on the T-Triangle due to the overlapping blockades, e.g., Q1 interaction radius blocks Q5 and Q2 interaction radius blocks Q6. Thus, it would take longer to run the operations on the T-Triangle topology.

On the other hand, a $CX$ gate on Q2-Q3 can be run directly on both the S-Triangle and T-Triangle topologies, but would require SWAP operations on the Square topology (e.g., swap Q2 with Q1 and then perform the $CX$ gate on Q1-Q3). Similarly, a two-qubit gate on Q2-Q6 can be run directly on T-Triangle, whereas the other two topologies would require SWAPs, thus, taking longer to run. These different trade-offs can affect different overall circuits differently. 

At the circuit level, evaluating the critical pulse count of different quantum benchmarks shows how certain circuits are better suited for different topologies. We show the critical pulse count performance of certain shapes on different well-studied quantum benchmarks in Table \ref{tab:difalgo} where each shape poses a unique advantage on different benchmarks. This can be seen in Table \ref{tab:difalgo}  where the critical pulse count evaluation has a different optimal topology for each benchmark shown. 

Depending on the execution metric (critical pulse count, total pulse count, and fidelity), the optimal topology can be different, even within the same circuit. We demonstrate this observation in Table \ref{tab:samealgo}. This is an important observation as it shows that it is not possible for a single model to maximize both objectives at the same time as they are at odds with one another. As each objective can has merit and can be justified given the specific requirements of an application for \sol{} opts for one model per objective so that the application of the model can dictate the model used.

Next, we demonstrate how \sol{} uses specific properties of a circuit to optimize for important objectives such as fidelity.

\section{Model Training}

A high-level overview of all of the steps of \sol{} is provided in Fig.~\ref{overview}. The design of \sol{} is split into two primary main sections: (1) model training, and (2) topology inference and circuit execution. As a summary, the first step is training a neural network on quantum descriptors and page rank feature analysis to learn optimal topologies for reducing pulse count along the critical path, which is described in this section. As described in the following section (Sec.~\ref{solution}), once this model is trained, it is ready to run inference to predict the optimal topology for a given circuit.  At this stage, the circuit is ready to be mapped to the topology by assigning logical qubits to physical hardware locations. Once mapping is complete, operations are scheduled and the circuit can be executed. 

\sol{} learns to anticipate the best topology for a set of quantum instructions to be performed. This model is trained by running multiple circuits with different parameters and learning the patterns of instruction sets that lend to unique best topologies for the total execution time of the circuit. Next, we describe how the predictive model is constructed and trained.

\subsection{Building the Model}

\sol{} analyzes many static features on our random circuits in order to train the model. These features aim to represent the eventual build of the circuit on different topologies without actually having to map and build the circuit. This process runs a page rank algorithm followed by a forward pass of a few small neural networks as opposed to an entire mapping, and routing procedure.

We use a variety of features in our neural networks, including quantum specific circuit descriptors and classical circuit descriptors. The idea is to understand at a high level how circuits that are similar and from there which types of circuits are best suited for different topologies. For example, if one circuit can be described as highly entangled and with a few number of unique qubits and is much more efficient on a square, then a similar circuit would more likely be most efficient if mapped to a square as well.

\subsection{Generating Training Data}
\begin{figure}[t]
	\centering
    \includegraphics[scale=0.18]{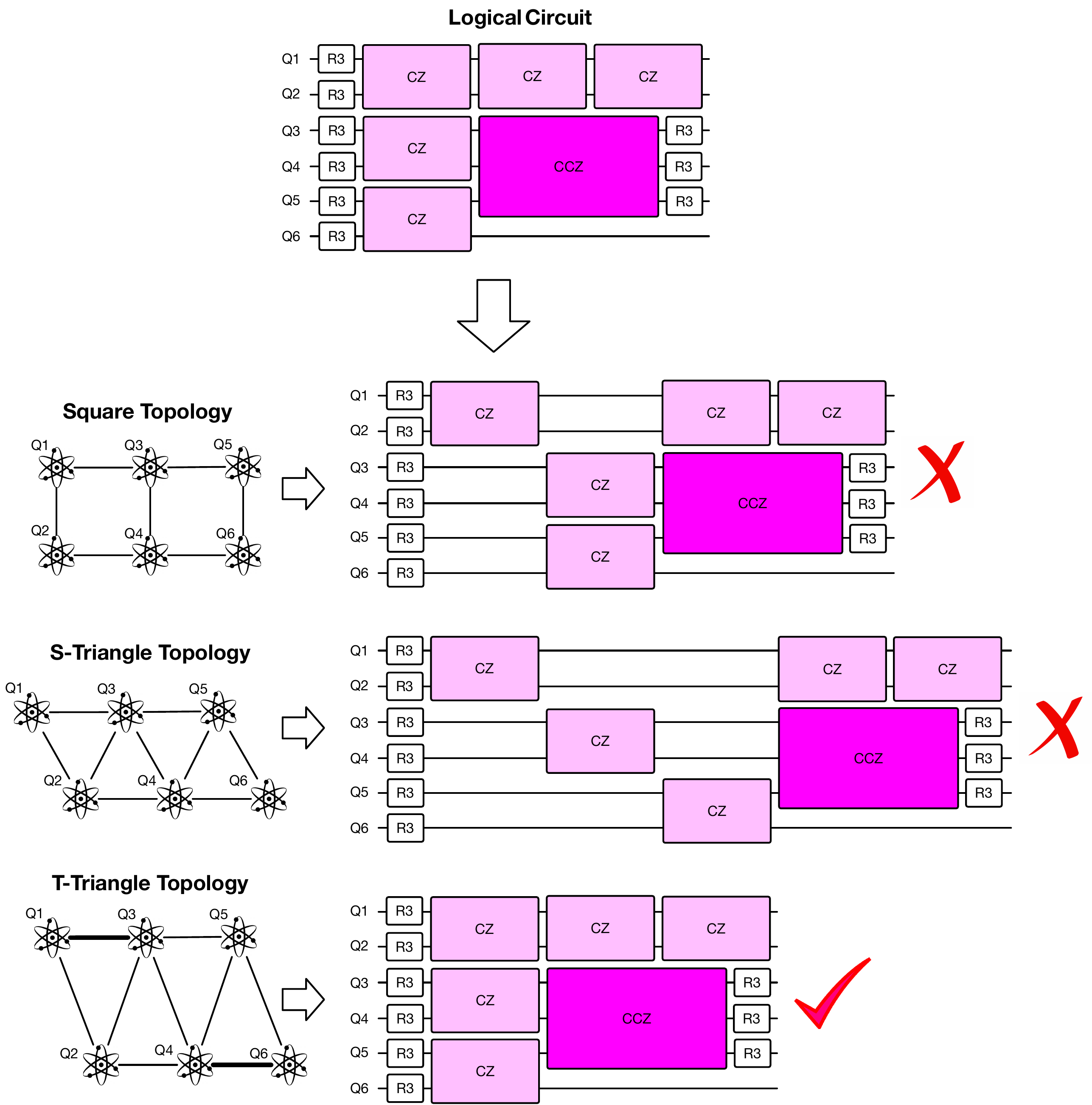}
    \vspace{1mm}
    \hrule
    \vspace{-3mm}
	\caption{The logical circuit passed into \sol{} is already parallelized. This transfer of the parallelism for neutral atom circuits is not trivial and must be reconstructed with a best-effort approach. However, some topologies are more suited to the parallelism of certain logical circuits than others.}
	\label{select}
	\vspace{-4mm}
\end{figure}

The ability to leverage different topologies is one of the most useful aspects neutral atom computing. \sol{} works to leverage this feature by identifying unique topologies to best best a circuit on. We identify the most likely best topology out of a set of topologies based on experimental analysis.  This experimental analysis is based on curated random instruction set generation and analyzing the effects of composing them on differing topologies. Investigating features in these instruction sets through the lens of a multivariate Gaussian generalizes well for selecting the best topology (Fig.~\ref{select}).

\subsubsection{Building Randomized Instruction Sets}
We generate realistic instruction sets which model real-world benchmarks.  The goal is to be able to generalize over these benchmarks based on training on similar benchmarks. A few parameters specify the composition of these instruction sets.  Firstly, we define inter-connectivity, whereby at each step of adding a new instruction, the probability to sample from previously used qubits as opposed to selecting a new one.  The higher the inter-connectivity, the lower the overall complexity of the circuit and likely the longer the critical path. The second parameter is the three qubit gate rate. The higher this rate, the more SWAP gates are typically necessary, as it is less likely for three gates to be near each other than 2. We also specify the one qubit gate rate as the rate in which one qubit gates are generated.  Additionally, the circuit width is specified which outlines the unique number of qubits in the circuit. Finally, the minimum number of instructions is set to give a rough level of instruction count to the circuit.  As \sol{} is designed to be executable on multiple topologies, these instruction sets are defined by the length of the critical path (without SWAPs as these change from one topology to another), the three qubit gate rate, the one qubit gate rate, and on the circuit width and are tested on all topologies.  These instruction sets are then analyzed using the feature analysis previously mentioned, followed by scheduling and execution to label the "ground truth" measurements for different topologies.  The goal of the neural networks is to match these regressions for new, unseen instruction sets.


\subsubsection{Quantum Descriptors used in Neural Network}

The circuit width, $W$, is the number of unique qubits, $q$ used within the circuit. This metric is useful for understanding the eventual mapping on neutral atom circuits as it dictates how many nodes need to be mapped. This number strongly affects the best topology for the circuit. Circuit width is defined formally as 
\begin{equation}
W=q_{active}
\end{equation}

Second, we consider the circuit depth, $D$, which is the longest critical path of the circuit and is defined as:
\begin{equation}
D=\sum_{j}^{t}{max(Q)}
\end{equation}
The critical path is the shortest path the circuit must take in order to complete.  Circuit Depth infers the limit to parallelism of the circuit.  This serial path cannot be parallelized and is a lower bound for the execution time of the circuit.

We then calculate the gate density, $\rho$ as
\begin{equation}
\rho = \dfrac{G_{1 qubit} + 2*G_{2 qubit} + 3*G_{3 qubit}}{W*D}
\end{equation}

where $G_{1 qubit}$ represents the total number of single qubit gates, $G_{2 qubit}$ represents the total number of 2 qubit gates, and $G_{3 qubit}$ represents the total number of 3 qubit gates. The gate density describes the level of occupancy of gates throughout the circuit.  This gives insight to the likely occupancy once the circuit is mapped to quantum hardware.

We use entanglement variance, $EV$ to encapsulate the level of interconnectedness of the graph, formalized as:
\begin{equation}
EV = \dfrac{\log{\sum_{i=0}^{W}{G_{qi}(2qubits) - \mean{G_{qi}(2qubits)}}}}{W}
\end{equation}

We use program communication, $PC$,  as the normalized average degree of the interaction graph \cite{SupermarQ} to understand the connection between pairs of qubits.
\begin{equation}
PC = \dfrac{\sum_{i=0}^{W}{d(q_i)}}{W(W-1)}
\end{equation}
where $d(q_i)$ represents degree of qubit $i$.

Critical Depth, $CD$,  measure the shortest possible duration the circuit must take in order to complete in terms of two-qubit gates. This is useful as it correlates highly to circuit fidelity \cite{SupermarQ}.  
\begin{equation}
CD = n_{e_d} / n_e
\end{equation}
where $n_{e_d}$ is a count of the two-qubit gates on the critical path and $n_e$ is a count of the total two-qubit gates

Entanglement Ratio, $ER$, is a rough measure of the amount of entanglement throughout the circuit. It is difficult to measure this throughout every point in the circuit as this would require access to the full state-vector \cite{SupermarQ}. Instead we measure the ratio of the count of all two-qubit gate operations $n_e$ to the count of all gate operations within the circuit $n_g$. 
\begin{equation}
ER = n_e / n_g
\end{equation}

\subsubsection{Leveraging Page Rank Features}

We also use features not specific to quantum circuits.  We leverage a few features from Page Rank \cite{page1999pagerank} in order to understand the variance in entanglements within a given circuit. Prior to this, there has been a large effort in using PageRank in quantum computing \cite{loke2017comparing, paparo2013quantum}.  We use the page rank variance of the circuit in order to understand the level of dependency between qubits. For this we start by calculating the weighted adjacency matrix of the circuit.  This results in a $W$x$W$ matrix with 0's along the main diagonal. We then run page rank on this matrix until the page rank loss is below a fixed $\epsilon$.  This adds rich data to our model in the form of page rank standard deviation, page rank max value, and page rank mean. 
After initialization, the Page Rank for each qubit $U$ is
\begin{equation}
PageRank(U) = \sum_V{\dfrac{PageRank(V)}{\abs{Out(V)}}}
\end{equation}

From the page rank solution vector, we take the Page Rank Mean, $\mu$ as 
\begin{equation}
\mu = \dfrac{\sum{PageRank(U)}}{W}
\end{equation}

We calculate Page Rank Standard Deviation, $\sigma$ as
\begin{equation}
\sigma = \sqrt{\dfrac{\sum\limits_{U \in W}{(U-\mu)}}{W}}
\end{equation}
The standard deviation of Page Rank provides insights regarding the distribution of central nodes within the circuit.  A high standard deviation would indicate for example that much of the complexity within the circuit resides within a fewer number of qubits as opposed to a low standard deviation which would imply otherwise.

The maximum Page Rank value, $M$, informs about the central dependencies of the circuit, where a high value indicates many swaps to interact with the center and less opportunity for parallelism. For  all qubits $U \in W$, the maximum value is defined as below and is bound by $(0,1]$.
\begin{equation}
M = max(PageRank(U)) 
\end{equation}

\subsubsection{Gate Size Proportions}

The final set of features we add is based on the distribution of gate types.  We calculate the proportion of 1-qubit gates, 2-qubit-gates, and 3-qubit gates within the circuit. These values are each bounded from [0,1] and all sum to 1.

\subsection{Training the Neural Network}

We train our model on randomly generated circuits aimed at emulating realistic quantum circuits. We train on instruction sets varying in size from 20 to 100 operations not including swap operations as these represent a variable amount of additional instructions calculated at run-time. We generate circuits with widths between 5 and 70. The core of this model is a small neural network that transforms these static circuit features to a label representing the optimal topology. In the case of \sol{}, only the tessellating shapes are considered, however; this neural network can be expanded to any number of shapes by tweaking the dimensionality of the output layer. 5 fold cross validation is used to split our training and testing data with 80\% used for training and 20\% used for testing.  

\subsubsection{Network Architecture}

Due to the small number of features, only a small network is needed for the classification task. The model is composed of the input layer of size 14, followed by two hidden layers each of size 15, and an output layer of size 1.  The input layer has one neuron for each feature to be able to encode all of the input information.  The hidden layers use relu activation functions to regularize the training.  The output layer is of size 1 to perform the regression. We use Adam Optimizer \cite{kingma2014adam} for training with mean absolute loss and train over 400 epochs. 

\subsubsection{Shape Specific Networks}

For each regression task of estimating the total pulse count, critical pulse count, and fidelity, \sol{} uses different sets of neural networks for regression. This ensures that each network has a single objective, allowing it to be smaller and easier to train. For each shape, a separate network is trained in order to learn the mappings from the feature space to each measurement (total pulse count, critical pulse count, and fidelity). The classification is then the shape corresponding to the network that produces the minimum value in the respective measurement. This significantly outperform classification based networks as the actual measurement are important for deciding the optimal shape.

\subsubsection{Hyperparameters}

All networks use identical hyperparameters.  We set the learning rate to $.001$. The constant for numerical stability, $\epsilon$, is set to $1e^{-7}$. The first order exponential decay variable, $\beta_1$, is set to $.9$.  Additionally, the second order exponential decay variable, $\beta_2$ is set to $.999$. 
\section{Topology Inference and Circuit Execution}
\label{solution}

\noindent\textbf{\sol{} uses a systematic approach driven based on graph analysis methods to represent the neutral atom circuits.} By taking into account familiar look-ahead properties, we factor in future circuit operations in building the circuit.  The stages required to perform topology inference and circuit execution follow a trained model that informs the optimal shape to be mapped.  From here the circuit is mapped to the specified topology, followed by a scheduling procedure to ensure operations do not interfere with one another, are computed in the proper order, and are scheduled as parallel as possible.  One the schedule is complete, the circuit has a timing of operations and is ready to be executed or simulated.  The overview of \sol{}'s approach can be found in Figure \ref{overview}.

\subsection{Mapping}
The first step \sol{} takes to orchestrate neutral atom circuits after prediction, is to map the logical qubits to physical locations on a 2-D lattice. While previous works have analyzed performance on a single topology such as a square grid, we build our approach to generalize well over multiple topologies (Fig.~\ref{topos}).

Regardless of the identified shape for the circuit, the first step of the mapping procedure is identical. Based on the analysis of the circuit structure, \sol{} first identifies the pair of qubits that are used most commonly together. This count includes qubits present together in operations greater than 2 qubits as well. Identify the center-most position within the grid and assign the first logical qubit to that position.  Find the nearest location adjacent to the center and assign the second logical qubit in the pair there. The first location $L_{0}$ is calculated over all points $L$ where
\begin{equation}
L_{0} = \operatorname*{argmin}_{p} \sum_{P}  (P-P_n)^2 
\end{equation}
This heuristic is based on the idea of reducing SWAP operations by keeping the commonly used qubits in the center and thus the most easily accessible overall by other qubits. Reducing SWAPs is especially important as they require many more pulses and therefore time to complete then other gates as well as compromising a lower output fidelity.

\textbf{The second stage of the mapping procedure is to build an adjacency map.} This adjacency map is comprised of edges as connections of all qubits within the circuit. Each pair of qubits present in the graph has a weight that defines the relationship between the two qubits. This relationship is defined and used to assign weights to each edge between all qubits $u$ and $v$ as the count of the connections between them. This compromises the weight function $w(u, v)$. 


From there, one at a time, qubits are assigned to the graph in order based on their weighted interaction to all other qubits currently existing on the graph $v$. This objective refreshes after each addition to the graph and must be recalculated with the inclusion of the new qubit.
\begin{equation}
Q_{next} = \operatorname*{argmax}_{q} \sum_{u}\sum_{v}  w(u,v)
\end{equation}
Once the new qubit to be added has been decided, the location is assigned to the new qubit according to a weighted sum $w(u,v)$ of the interaction strength and the inverse distance between the selected qubit $u$ and each other qubit $v$ where each possible location within the space of (x, y) where (x, y) is previously unassigned.
\begin{equation}
L = \operatorname*{argmax}_{(i, j) \in (\prime{x}, \prime{y})} \sum_{v} \frac{w(u,v)}  {\sqrt{(i-v_i)^2 + (j - v_j)^2}}
\end{equation}

In the case of a 3 dimensional lattice with an additional $z$ axis, this objective presents itself as follows.
\begin{equation}
L = \operatorname*{argmax}_{(i, j, k) \in (\prime{x}, \prime{y}, \prime{z})} \sum_{v}  \frac{w(u,v)} {\sqrt{(i-v_i)^2 + (j - v_j)^2 + (k-v_k)^2}}
\end{equation}
After this procedure, the mapping is complete and the process repeats until all qubits have an assigned.  


\subsubsection{Inserting SWAP Gates}
Once all qubits have a physical assigned location, it may be necessary to add SWAP operations in order to interact qubits that are too far from one another for a Rydberg interaction to entangle them directly. It is important to note that it is possible for any given operation to require multiple SWAPs to complete. For each instruction in the circuit's instruction set, the locations of the qubits are checked to see if they are all within the effective Rydberg interaction radius for the gate type. If it is the case that these qubits do not all fall within a given interaction radius, then SWAPs are added in a best effort shortest path to connect all of the qubits.

In the case of 2 qubits, this procedure seeks to find the shortest path by adding SWAPs one at a time going from point $u$ to $v$. This is done by swapping $u$ with a new point $h$ that then becomes the new $u$. Every location $u_l$ that is within the 2-qubit Rydberg interaction radius is considered to become the new $h$.
\begin{equation}
h = \operatorname*{argmin}_l{(u_{l} - v)^2}
\end{equation}

Once all qubits are within the necessary interaction radius, all recorded SWAPs are added to the schedule where each SWAP's previous SWAPs are noted as previous requirements.  


\subsection{Scheduling}
Once the SWAPs are added, the operations are ready to be scheduled. The goal of this stage is to preserve the necessary order of logical operations while maximizing the operations that can be run in parallel at the same time. Scheduling properly is critical in ensuring well-parallelized circuits that fully take advantage of the given topology in the neutral atom circuit. As this procedure is designed to be neutral atom hardware agnostic, we schedule with abstract time steps.  With this, each gate operation i.e., (SWAP, $CX$, $CCX$) are assigned a specified amount of time steps for completion that which are defined by the number of pulses required in order to respectively complete each gate operation. 

\subsubsection{Scheduling: Creating the DAG}

The first step in organizing the circuit operations into a Directed Acyclic Graph is to give all instructions a unique identifier.  Once this is complete, it is necessary to build a dependency graph where operations that rely on a specific qubit ensure that the qubit is placed in the proper state by prior operations before scheduling the newer operation.  On the same note, all operations that can be parallelized in a given time step are made parallel. It is also important to note that the interaction radius for the specific gates in addition to the time each gate takes to complete are necessary parameters for creating the DAG. \sol{} uses a greedy approach in constructing the DAG where it looks for all operations that can be completed at a given timestep, schedules them, then advanced one time step and checks again. Once the DAG is compiled, nothing else is needed for execution to commence. Since commercial neutral atom machines are not yet publicly available, we build a circuit identical to the DAG in Qiskit.  This Qiskit simulation allows us to add realistic, synthetic error and enables comparison of our models across the DAG created from multiple topologies.


\section{Evaluation and Results}

\begin{figure}[t]
	\centering
    \includegraphics[scale=.55]{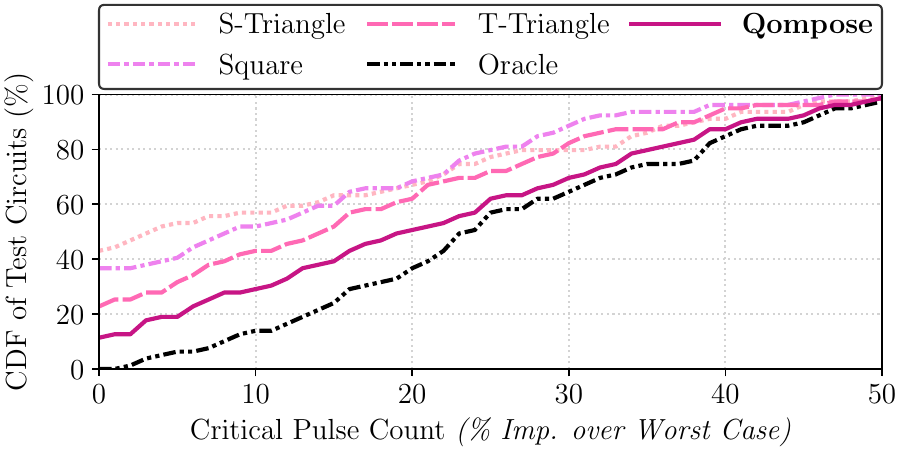}
    \vspace{1mm}
    \hrule
    \vspace{-3mm}
	\caption{Critical pulse count CDF on randomly generated circuits.  The lower the line, the better, as it represents the most improvement over the worst case. Oracle is computed to find the ground truth best topology for the lowest pulse count along the critical path, where the worst case represents the ground truth worst case. The individual shapes represent the method of selecting a single shape for all cases.}
	\label{critical_path_cdf}
	\vspace{-4mm}
\end{figure}

\sol{} demonstrates highly accurate predictions on diverse random circuits as well as on established benchmarks.  We measure performance by the number of pulses required for total critical path as well as the total pulse count and the overall circuit fidelity. These measurements ensure decreased error, lower execution time, and smaller probability of a necessary circuit reset. 

The benchmarks we include are Walk-Search \cite{shenvi2003quantum}, Decoder \cite{parent2016compilation}, Mod5d1 \cite{parent2016compilation}, Bernstein-Vazirani \cite{bernstein1997quantum}, Multiply-10 \cite{obenland1998parallel}, Simon's Algorithm \cite{simon1997power}, QAOA \cite{farhi2014quantum}, VQE \cite{peruzzo2014variational}, HLF, 32bitAdder \cite{parent2016compilation} , CH \cite{fips1995180}, XY \cite{obenland1998parallel} \cite{abrams2020implementation}, TFIM \cite{tomesh2022supermarq}, BitwiseMajority32 \cite{parent2016compilation}, ISING \cite{li2020qasmbench}, Mult15 \cite{obenland1998parallel}, SECA \cite{li2020qasmbench}, QFT15 \cite{li2020qasmbench}, and MULT25 \cite{obenland1998parallel}. These benchmarks are selected to span circuits of various sizes and diverse distributions of one, two, and three-qubit gates. As some benchmarks consist of a large number of qubits, they are not feasible to run error analysis on due to the impossibility to perform ideal classical simulation. We therefore split these benchmarks into small and large circuits where we can run fidelity analysis on the smaller circuits.

\subsection{Minimizing Pulse Count for Critical Path}

Here we demonstrate how \sol{} is effective at determining effective topologies for reducing the number of pulses in the critical path of a circuit. The pulse count in the critical path is important to neutral atom circuits for a few main reasons.  First, the circuit fidelity is tightly coupled with this path as the longer the critical path is, the more error can compound due to gate error and decoherence effects. In the NISQ era of quantum computing, it is crucial to work around the high level of noise, and by optimizing for the length of the critical path, one is able to mitigate the effects of compounding noise. Additionally, overall execution duration is proportional to this length, so by reducing it, executions will be faster.  In Figure  \ref{critical_path_cdf}, we demonstrate how \sol{} is the best overall at predicting the best topology for a given instruction set on our generated circuits and is in fact quite close to the Oracle. \sol{} is 19\% faster over the worst case overall.

\subsubsection{Critical Path Benchmarks}

\begin{figure}[t]
	\centering
    \includegraphics[scale=0.41]{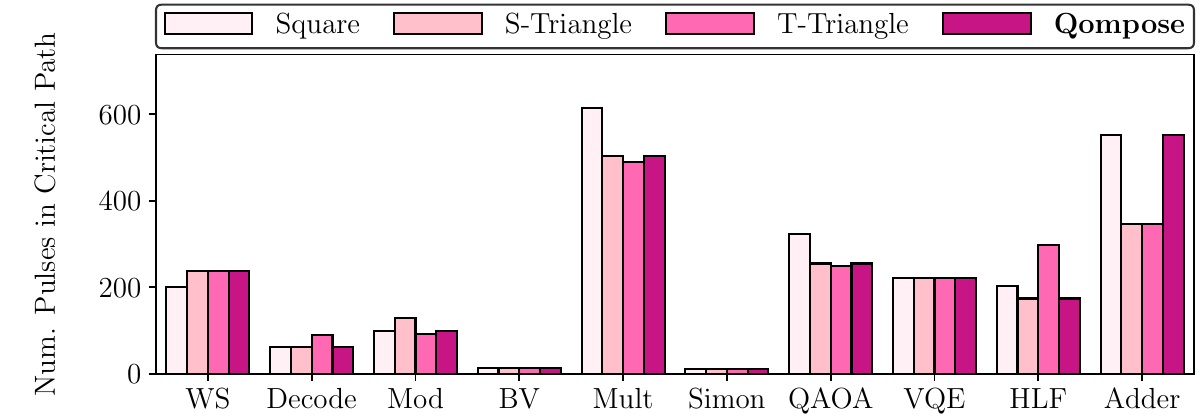}
    \includegraphics[scale=0.41]{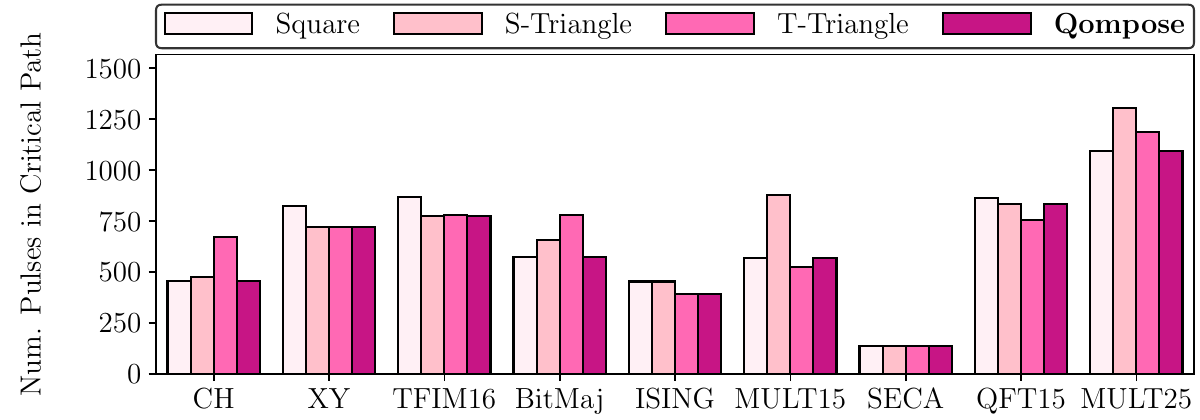}
    \vspace{1mm}
    \hrule
    \vspace{-3mm}
	\caption{Critical pulse count on quantum benchmarks.  Some benchmarks are better suited for square topology, some for s-triangle, and some for t-triangle. \sol{} selects the optimal topology in most instances, and comes close in most cases where it does not select the best one.}
	\label{critical_benchmarks}
	\vspace{-3mm}
\end{figure}

In Figure \ref{critical_benchmarks}, we explore how \sol{} performs on real-world quantum circuit benchmarks.  These benchmarks vary significantly in the circuit descriptors we analyze, which is intended as to understand how \sol{} generalizes to different types of unique circuits. We find that each shape poses unique advantages for different benchmarks, and therefore no single shape is the best solution for all cases. Selecting the proper shape is important for minimizing the critical path.  In CH, a function used in SHA-256, requires 32.3\% less pulses to run on a t-triangle topology as compared to a s-triangle one.

\subsubsection{Evaluating Critical Path Feature Importance}
We evaluate the importance of each feature by setting them to their average value before performing the regression again with the altered input. This technique is an attempt to neutralize each given feature as to understand the impact on regression without it. The features are ranked by the loss incurred by swapping the values with the mean.  We demonstrate this in Table \ref{tab:ranking}.

\begin{table}[t]
\caption{Neural network feature importance ranking}
\small{
\vspace{-3mm}
\begin{tabular}{|| c c | c c ||} 
\hline
Rank & Feature & Rank & Feature\\ [0.5ex] 
\hline\hline
1 & Num. Instructions & 8 &  Gate Density\\ 
\hline
2 & Circuit Width & 9 &  Critical Path\\
\hline
3 & Program Comm. & 10 &  Two Qubit Gates\\
\hline
4 & Entanglement Variance & 11 &  Page Rank Variance\\
\hline
5 & One Qubit Gate Prop & 12 &  Entanglement Ratio\\ 
\hline
6 &  Page Rank Max & 13 &  Circuit Depth\\
\hline
7 &  Page Rank Mean & 14 &  Three Qubit Gates\\
\hline
\end{tabular}
\label{tab:ranking}
}
\end{table}

\subsection{Minimizing the Total Pulse Count}

While the critical pulse count is important to understanding elements of circuit fidelity and total circuit duration time, total pulse count also measures important aspects of neutral atom circuits.  Every time a pulse occurs in a neutral atom circuit, there is a some probability that the atom is lost and the entire circuit needs to be recompiled and rerun. In this section we analyze how \sol{} improves the total pulse count over randomly generated circuits and on a series of real-world benchmarks.

\begin{figure}[t]
	\centering
    \includegraphics[scale=.55]{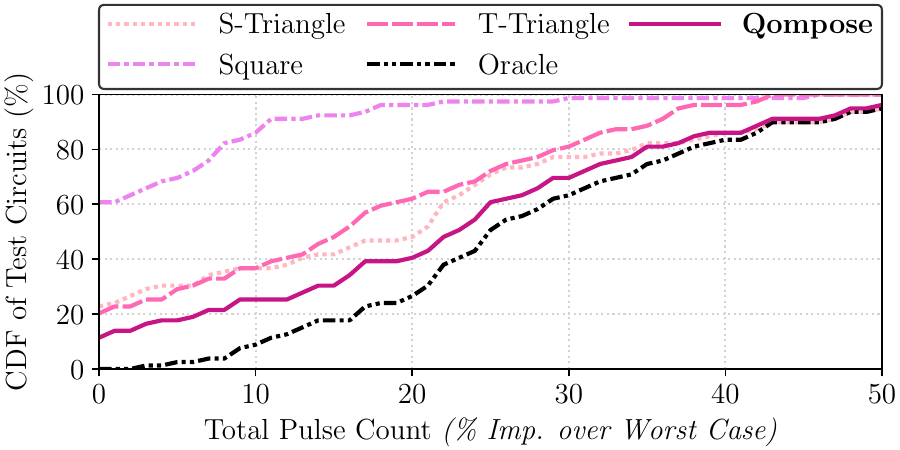}
    \vspace{1mm}
    \hrule
    \vspace{-3mm}
	\caption{Total pulse count of randomly generated circuits.  Overall \sol{} is close to the Oracle and significantly improves over the worst case. \sol{} is also better than selecting any given shape to optimize for lower total pulse count.}
	\label{total_path}
	\vspace{-3mm}
\end{figure}

\subsubsection{Total Pulse Benchmarks}

\begin{figure}[t]
	\centering
    \includegraphics[scale=0.41]{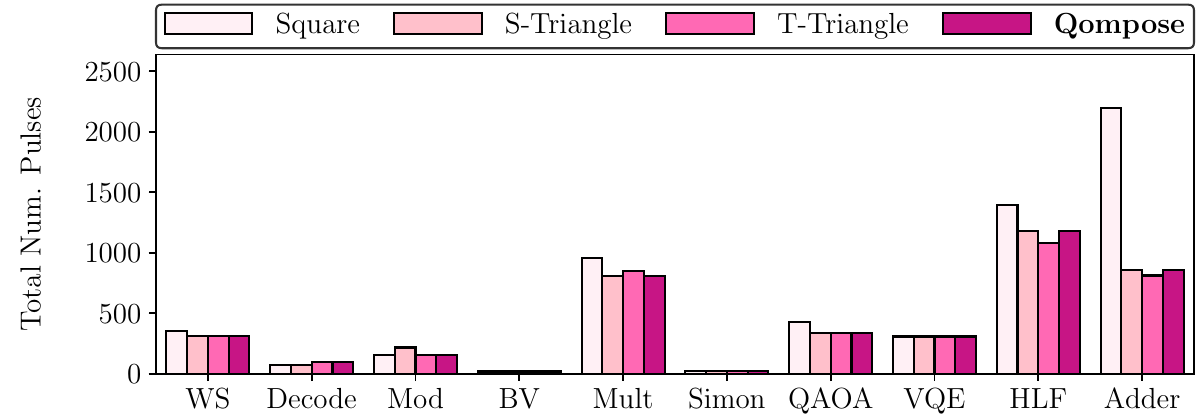}
    \includegraphics[scale=0.41]{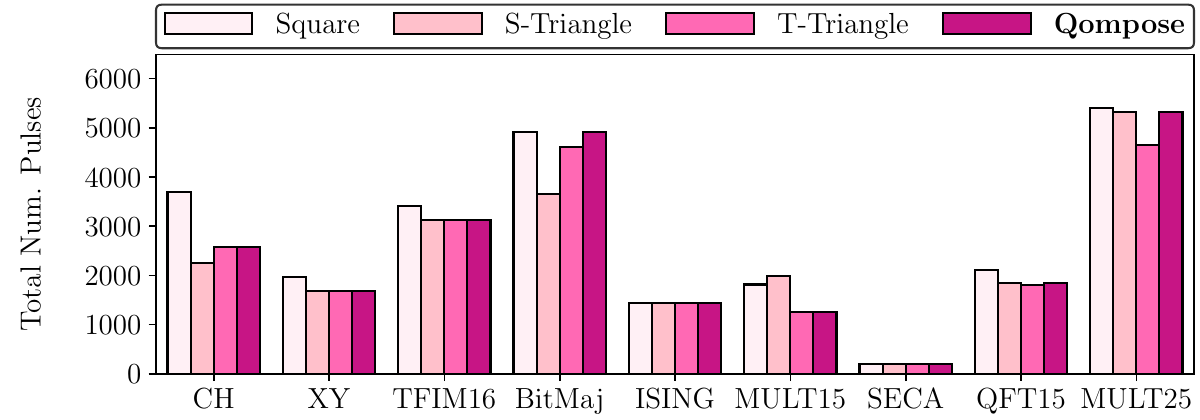}
    \vspace{1mm}
    \hrule
    \vspace{-3mm}
	\caption{Total pulse count of benchmarks: \sol{} accurately predicts the optimal topology to minimize the total pulse count on established benchmarks, which is not always the same as the one for reducing the critical pulse count.}
	\label{total_path_benchmarks}
	\vspace{-4mm}
\end{figure}

In Figure \ref{total_path_benchmarks}, we investigate the CDF of the total pulse count on randomly generated circuits. \sol{} remains the best decision procedure and is tightly coupled to the oracle. For total pulse count, the square topology is the worst overall.  This is an interesting finding, especially considering this topology being the de facto one used in neutral atom computing.  If the probability a circuit must be reset is high, this would suggest to opt against using the square topology as to avoid this costly procedure.

\begin{figure}[t]
	\centering
    \includegraphics[scale=0.41]{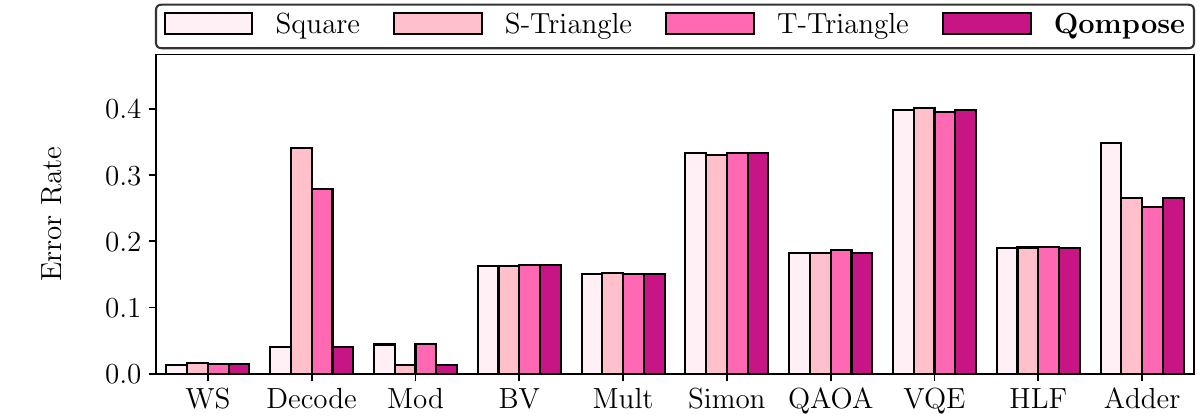}
    \vspace{1mm}
    \hrule
    \vspace{-3mm}
	\caption{Error rate for real-world quantum benchmarks. As simulation becomes exponentially more difficult with higher numbers of qubits, we test error on our first set of smaller benchmarks circuits.}
	\label{error}
	\vspace{-4mm}
\end{figure}

\subsection{Circuit Fidelity}

Error rate is an important measure during the NISQ era of quantum computing. Large error rates inherent in quantum circuits consist of state preparation and measurement (SPAM) errors, gate operation errors, and amplitude and phase dampening errors all compound for large error overall. We analyze the bitwise error both on random circuits and established benchmark circuits in Figure \ref{error}. For experimental evaluation, we use a noise level of $.01$ for one-qubit gates, and tesnor this model with itself for two-qubit gates, and once more for three-qubit gates. Each item of training and inference data is generated using the Qiskit Aer simulator with 5000 shots through the circuit.  The high amount of shots is used to gain a representative probability distribution of the expected error rates.

\section{Related Work and Conclusion}

In this work, we present \sol{}, a technique to select optimal algorithm-
specific layout for neutral atom quantum architectures. We provide an end-to-end procedure for deciding on topology, mapping, scheduling, and execution for generalized quantum circuits on neutral atom quantum computers. The foundation for this work has been laid by \cite{baker2021exploiting} by providing a basis for mapping and routing for neutral atom circuits. However, this work did not explore the design space of optimizing the atom topology; it only explored the widely-used Square topology. In contrast, \sol{} shows that other topologies can far outperform the Square topology for certain quantum algorithms.

Other works have provided thorough review into neutral atom quantum computing \cite{weiss2017quantum, henriet2020quantum,saffman2016quantum} -- they explain the procedures and detail the experimental parameters that we build our simulator on. As a note, circuit mapping and scheduling works such as \cite{li2019tackling,patel2021qraft} have tackled the logical-to-physical execution problem for qubits; however, all of these works have focused on the space of superconducting hardware and not on neutral atom hardware.

On the other hand, in \sol{}, we utilize many quantum and classical circuit descriptors that are leveraged by neural networks to estimate the fidelity, critical pulse count, and total pulse count for each shape, then select the best one. In addition we demonstrate specific advantages of different topologies for different circuits based on fidelity, critical pulse count, and total pulse count. We demonstrate that \sol{} reduces the critical pulse count over randomly selecting a topology by 8.4\% for randomly generated circuits and 5.4\% for real-world benchmarks. \sol{} is open-sourced for community-wide experimentation and adoption: \url{https://anonymous.4open.science/r/Qompose-5B67/} 

\bibliographystyle{ACM-Reference-Format}
\newpage
\bibliography{main}

\end{document}